# A Reinforcement Learning Approach for Intelligent Traffic Signal Control at Urban Intersections

Mengyu Guo, Pin Wang*, *Member, IEEE,* Ching-Yao Chan, *Member, IEEE* and Sid Askary

*Abstract*— Ineffective and inflexible traffic signal control at urban intersections can often lead to bottlenecks in traffic flows and cause congestion, delay, and environmental problems. How to manage traffic smartly by intelligent signal control is a significant challenge in urban traffic management. With recent advances in machine learning, especially reinforcement learning (RL), traffic signal control using advanced machine learning techniques represents a promising solution to tackle this problem. In this paper, we propose a RL approach for traffic signal control at urban intersections. Specifically, we use neural networks as Q-function approximator (a.k.a. Q-network) to deal with the complex traffic signal control problem where the state space is large and the action space can be discrete. The state space is defined based on real-time traffic information, i.e. vehicle position, direction and speed. The action space includes various traffic signal phases which are critical in generating a reasonable and realistic control mechanism, given the prominent spatial-temporal characteristics of urban traffic. In the simulation experiment, we use SUMO, an open source traffic simulator, to construct realistic urban intersection settings. Moreover, we use different traffic patterns, such as major/minor road traffic, through/left-turn lane traffic, tidal traffic, and varying demand traffic, to train a generalized traffic signal control model that can be adapted to various traffic conditions. The simulation results demonstrate the convergence and generalization performance of our RL approach as well as its significant benefits in terms of queue length and wait time over several benchmarking methods in traffic signal control.

## I. Introduction

Traffic signal control has been a long-standing topic in urban traffic control. Effective traffic signal control, plays a significant role in alleviating traffic congestion, reducing waiting times of travelers, improving throughput of a road network, lowering vehicle emission and fuel consumption [1], [2]. However, optimal control of traffic signals at urban road intersections suffers from challenges associated with spatial-temporal dynamics of urban traffic. In the temporal dimension, urban traffic exhibits significant variations in peak and non-peak hours. For example, in large urban areas like Beijing, nearly 40% of daily traffic happens during peak hours (i.e. 7:00-9:00 and 17:00-19:00) [3]. Moreover, the variations in urban traffic also exist in the spatial dimension. For example, tidal traffic pattern (i.e. traffic demand of one direction is higher than the opposite direction) has long been recognized as an important spatial feature of peak hour traffic [4]. Similarly, non-uniform traffic demands are also common in the cases of major and minor roads, left-turn and through lanes [5]. Therefore, in order to cope with the spatial-temporal characteristics of urban traffic, traffic signals should be intelligently adaptive to real-time traffic demands.

Conventional methods for traffic signal control can be categorized into two classes: fixed-time control [6] and adaptive control [1]. Fixed-time control pre-calculates signal phase durations based on estimated traffic demands but often performs poorly when dealing with real-time traffic dynamics. Adaptive control addresses this shortcoming by dynamically adapting signal phase durations based on real-time traffic data from sensors (e.g. camera, radar). However, difficulties still exist: (1) extensive field-tests and manual tuning are required to reflect roadway and traffic characteristics; the process is resource-consuming and the outcomes cannot be ported generally; (2) given the complexity of urban traffic systems, it is often not feasible to develop an accurate model-based method without simplified assumptions about operating conditions - which can lead to suboptimal solutions [7], [8].

With increasing availability of infrastructure sensors and crowd-sourced data, it is now possible to collect large volumes of real-time traffic data from operations of urban road networks [1], [2]. Abundant real-time traffic data together with recent advancements in computational methods has the potential to make significant improvements in traffic signal control. In particular, the successful applications of reinforcement learning (RL) in solving complex problems suggest a promising opportunity to endow traffic signals with a higher degree of intelligence. A number of previous studies based on such machine learning concepts have been pursued along this line of thought [9]-[16]. However, to the best of our knowledge, the existing research does not address the complex dynamics of urban traffic, mainly because the proposed methods are often trained and tested in a single traffic pattern. Thus, the control policy learned from a limited training set cannot be generalized to different traffic patterns and its effectiveness cannot be fully evaluated.

In this paper, we extend the application of reinforcement learning for traffic signal control in two aspects: firstly, a Q-learning algorithm enhanced with a custom neural network-based Q-function approximator is developed to enable traffic signal control with both *real-time* and *high-dimensional* traffic information, such as vehicle position, direction and speed. Various practical signal phases are included in the action space in order to generate a reasonable and realistic control mechanism - given the prominent spatial-temporal characteristics of urban traffic. Moreover, inspired by recent studies in RL [17], several state-of-the-art techniques such as *experience replay* and *target network* are employed to facilitate a fast convergence

Mengyu Guo, Pin Wang (*corresponding author) and Ching-Yao Chan are with California PATH, University of California, Berkeley, Richmond, CA 94804 USA (e-mail: {mengyu, pin_wang, cychan}@berkeley.edu).

Sid Askary is with Futurewei Technologies, Inc. Santa Clara, CA 95050 USA (e-mail: sid.askary@huawei.com).

of our approach. Additionally, four typical urban traffic patterns, namely: (1) *major/minor road* traffic, (2) *through/left-turn lane* traffic, (3) *tidal* traffic, and (4) *varying demand* traffic are used to train and validate the generalization performance of our RL approach. To reflect a real-world urban intersection setting, all training and testing experiments were conducted in SUMO (Simulation of Urban MObility), which is a widely used, high-fidelity urban traffic simulator with flexible APIs (i.e. TraCI, Traffic Control Interface) for traffic signal control [18].

The reminder of this paper is organized as follows: Section II presents a literature review of related work. Section III presents the detail of our RL approach for traffic light control. Section IV describes the simulation experiments and presents the experimental results. Section V concludes this paper and discusses directions for future research.

## II. RELATED WORK

In the past two decades, RL based traffic signal control has attracted significant attention from both academia and industry - with a growing body of work on this subject. El-Tantawy et al. [10] summarized the works from 1997 to 2010 that used RL to control traffic signal timing. During this period, the RL techniques were limited to tabular Q-learning and a discrete state space was normally used in small scales. For example, the queue length of an incoming lane was often categorized into "low", "medium" and "high" levels based on hand-crafted rules [11]. The complexity of the traffic environment at a traffic intersection could not be fully represented by such limited information. When more useful and relevant information was omitted from the limited states, the techniques seemed unable to act optimally in traffic signal control [16].

Enhanced with the remarkable neural network technology for function approximation, RL can now accomplish more difficult and complex tasks. Li et al. [13] proposed a deep Q-network with stacked auto-encoders (SAE) to approximate Q-function. The algorithm used the number of queued vehicles as states and trained with Q-learning algorithm to learn an optimal control policy at a single intersection. Shabestary et al. [14] used deep neural networks to operate directly on detailed sensory inputs and fed it into a continuous RL agent to optimize traffic signal control at signalized intersections. Zeng et al. [15] treated the traffic environment at a single intersection as a non-stationary environment and attempted to solve the traffic signal control problem with a memory-based RL method. Specifically, a recurrent neural network (RNN) was proposed to handle the sequence of observations and to approximate the optimal action value function. Liang et al. [16] quantified the complex traffic pattern as states by dividing the whole intersection into small grids. A convolutional neural network (CNN) was proposed to match the states and expected future rewards.

In summary, one limitation in previous work is the lack of consideration for the spatial-temporal characteristics of urban traffic in training and testing proposed methods. One of the main advantages of neural network is the capability of feature extraction from high-dimensional data. However, this aspect is not well-addressed in previous work. In this paper, the spatial-temporal characteristics of urban traffic is explicitly considered in various traffic patterns and the generalization ability of our RL approach is demonstrated.

## III. PROPOSED METHOD

### A. Problem Statement

In this paper, we consider a typical urban road intersection controlled by traffic signal. Pedestrian movements are not explicitly considered because in practice pedestrian movements are often served concurrently with accompanying through-vehicular movement without exclusive pedestrian phasing. As illustrated in Fig. 1, at the intersection there is one left-turn lane, one through lane, one shared through and right-turn lane in each incoming direction. Eight traffic signal phases in total are available; each corresponding to the right-of-way (i.e. green light) time interval assigned to a set of non-conflicting vehicular movements (see Fig. 2). In a scheme of fixed-time traffic signal control, the signal phases are visited sequentially in a pre-determined manner (i.e. fixed duration and sequence) regardless of traffic conditions. With the concept of intelligent traffic signal control, phases are visited in a more adaptive manner according to real-time traffic conditions. The system is aimed at guiding vehicular movements through the intersection more efficiently by minimizing queue length and wait time by maximizing throughput. This control process can be formulated as a reinforcement learning (RL) problem [12].

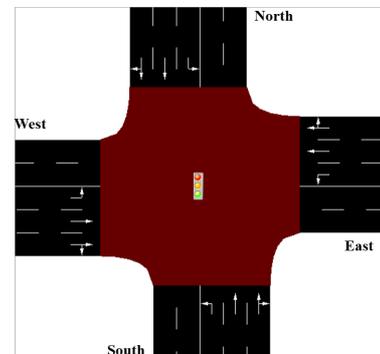

Figure 1. Layout of the signalized intersection under study.

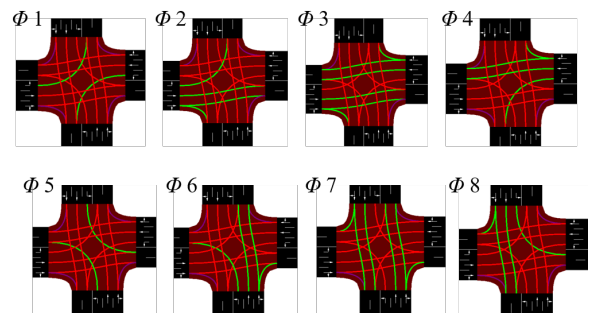

Figure 2. Eight signal phases at the intersection.

### B. RL Formulation

In RL, during training, there are three important elements ⟨$S$, $A$, $R$⟩, where $S$ is the state space ($s \in S$), $A$ is the action space ($a \in A$), and $R$ is the reward function. In the case of traffic signal control, we define these elements as follows:

*State*. Recent development in sensory technology makes real-time, vehicle-level traffic monitoring possible [1], [2]. Real-time traffic information such as individual vehicle's position and speed can be used as the states for RL-based traffic signal control [15], [16]. Here, we define the states based on two pieces of information, position and speed of vehicles as they approach an intersection. Specifically, we represent the state of an incoming lane as queue length $q$, which is defined as the real-time number of halting vehicles (i.e. vehicles with speed ≤ 0.1 m/s) on that lane. Hence, the state of the intersection in Fig. 1 can be characterized by a 12-dimensional vector $S = [q_1, q_2, …, q_{12}]$, corresponding to the 12 incoming lanes of the intersection.

*Action*. In order to maximize the efficiency of the intersection, both phase and timing of traffic signals are to be adaptively adjusted according to real-time traffic. Regarding the timing of traffic signals, an important decision is whether the current signal phase should be extended for a pre-defined time span to allow more vehicles to pass in this phase. As for the phase of traffic signals, it involves the decision of skipping certain phases according to real-time traffic conditions (i.e. no detection of approaching traffic), as implemented in fully-actuated control systems in real world. To that end, we define the action space as $A = \{\Phi_1, \Phi_2, …, \Phi_8\}$, corresponding to the eight signal phases at the intersection (see Fig. 2). Each chosen action denotes the current actuated signal phase. If the same signal phase is actuated consecutively, the duration of the phase will be extended for a time span $\Delta$. If a different signal phase is actuated, the traffic signals will transition to the next actuated phase after a pre-defined yellow signal duration.

*Reward*. The role of rewards in RL is to provide immediate feedbacks on the performance of previously chosen actions. In the setting of traffic signal control, performance criteria such as queue length and wait times are commonly used [14]. In accordance with our state definition above, we measure the total queue length of incoming lanes at training step $t$,

$$L_t = \sum_{i=1}^{12} q_i, \qquad (1)$$

and use

$$R_t = L_t - L_{t+1}, \qquad (2)$$

as reward at training step $t$. Hence, when the number of halting vehicles decreases between training step $t$ and $t+1$, a positive reward will be earned to encourage the chosen action.

### C. Q-network

Different from model-based control methods where full information of environment is required [7], [8], RL can learn the optimal strategy to control traffic signals via pure interaction with traffic environment. In this paper, we adopt Q-network in our RL approach. With a well-designed neural network as Q-function approximator, a Q-network could extract discriminative information from state space in order to derive an optimal action policy, thus avoiding the curse of dimensionality as the state-action space becomes huge [17], [20].

The goal of the Q-network is to train the traffic signal to adapt its phase and phase duration based on real-time traffic patterns at the intersection. This can be achieved by selecting an action at each training step ($a_t \in A$) that maximizes the expectation of accumulated future rewards (i.e. minimization of total queue length),

$$Q^*(s_t, a_t) = \max_\pi E[R_t + \gamma R_{t+1} + \gamma^2 R_{t+2} + … | \pi], \qquad (3)$$

where $\gamma \in (0,1)$ is the discount factor which represents the trade-off between future and immediate rewards. Current policy $\pi: S \times A \rightarrow [0,1]$ is defined as the probability of taking action $a_t$ in state $s_t$. According to the theory of dynamic programming [21], the optimal Q-function in (3) can be rewritten in the form of Bellman equation, assuming that in each training step the agent selects the action with the highest Q-value,

$$Q^*(s_t, a_t) = E[R_t + \gamma \max_{a'} Q^*(s_{t+1}, a')]. \qquad (4)$$

Traditional Q-learning algorithms solve the Bellman equation (4) in an iterative manner that rely on the concept "Q-table", which is a discrete table of Q-values [21]. However, given the definition of state space (i.e. queue length in vehicle number) in our formulation of traffic signal control, traditional Q-learning will suffer from the curse of dimensionality as the state-action space becomes huge. In our formulation, there would be $8 \times q^{12}$ Q-values (i.e. 8 actions and 12 dimensions of the state space). Even with a small number for queue length (e.g. $q = 5$), the total number of Q-values would still be astronomical (e.g. $\approx 2 \times 10^9$). Thus, in this paper we adopt the recent prominent idea of Q-network, i.e. using a neural network to approximate the Q-function:

$$Q(s, a : \theta) \approx Q^*(s, a), \qquad (5)$$

where $\theta$ represents the parameters of a neural network. Our Q-network for traffic signal control is illustrated in Fig. 3.

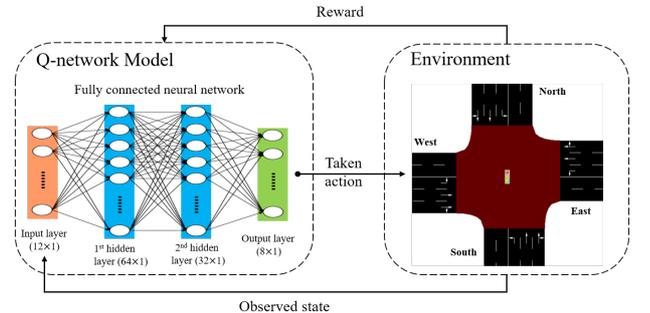

Figure 3. The architecture of the Q-network.

At the beginning of each training step, state information is collected and used as input layer to the neural network, then a feedforward process is applied which results in Q-value estimates at the output layer of the neural network. This neural network is optimized by the stochastic gradient descent algorithm (SGD) [21] and the loss function of the Q-network is defined as the Huber loss function [24]:

$$J(\theta) = E[z_t], \quad (6)$$

where $z_t = 0.5(y_t - x_t)^2$ if $|y_t - x_t| < 1$, or $z_t = |y_t - x_t| - 0.5$ otherwise. And $x_t$, $y_t$ are given by:

$$x_t = Q(s_t, a_t; \theta), \quad (7)$$

$$y_t = R_t + \gamma Q(s_{t+1}, \arg\max_{a'}(Q(s_{t+1}, a'; \theta)); \theta^-), \quad (8)$$

where $y_t$ is called the learning target of the network. The goal of this learning process is to obtain a set of parameters that facilitate accurate Q-value approximation.

Applying non-linear function approximators, such as a neural network, with a model-free Q-learning algorithm has some convergence problems [17]. Thus, two state-of-the-art techniques; *experience replay* and *target network* are employed in our Q-network to increase its stability during learning. Specifically, a memory *m* of recent experiences ⟨$s_t$, $a_t$, $R_t$, $s_{t+1}$⟩ are maintained and a batch of experiences is randomly sampled from the memory as training data at each training step. With experience replay, it is possible to break the correlation between successive samples [19]. We also use another neural network (i.e. target network) to compute the learning target (i.e. $y_t$ in (8)) and update the parameters $\theta^-$ after every $\Delta T$ training steps. This is helpful in stabilizing the target network during learning.

## IV. SIMULATION EXPERIMENT

In this section, we present the simulation experiment which includes: (1) SUMO simulation environment, (2) hyper-parameters of our Q-network, (3) traffic patterns at the intersection, (4) traditional traffic signal control methods used as benchmarks and (5) simulation results.

### A. SUMO Simulation Environment

As illustrated in Fig. 1, the signalized intersection is implemented in SUMO. The detailed settings of SUMO simulation environment are listed in TABLE I.

TABLE I. SUMO SIMULATION SETTINGS

| Parameter | Value |
| --- | --- |
| Lane length | 150 meters |
| Vehicle length | 5 meters |
| Minimal gap between vehicles | 2.5 meters |
| Car-following model | Krauss following model [22] |
| Max vehicle speed | 13.42 m/s |
| Acceleration ability of vehicles | 2.6 m/s$^2$ |
| Deceleration ability of vehicles | 4.5 m/s$^2$ |
| Duration of yellow signal | 3 seconds |
| Time span $\Delta$ of signal phase | 10 seconds |

### B. Hyper-parameters of the Q-network

Our Q-network is implemented with PyTorch [23] and communicated with SUMO simulation environment via TraCI (i.e. traffic control interface of SUMO). Both the SUMO simulations and the reinforcement learning process are run on a workstation with Intel Core i7-5930K CPU, 62.8 GB RAM and 1 Nvidia GeForce Titan X GPU. In all experiments, the Q-network is trained for 200 episodes, and each episode is a complete SUMO simulation with 1800 simulation steps. The hyper-parameters of our Q-network are listed in TABLE II.

TABLE II. HYPER-PARAMETERS IN THE Q-NETWORK

| Hyper-parameter | Value |
| --- | --- |
| Replay memory size $M$ | 10000 |
| Minibatch size $B$ | 128 |
| Starting $\varepsilon$ | 0.9 |
| Ending $\varepsilon$ | 0.05 |
| Target network update interval $\Delta T$ | 1800 simulation steps (i.e. one episode) |
| Discount factor $\gamma$ | 0.999 |
| Learning rate (SGD) | 0.01 |
| Momentum (SGD) | 0.9 |

### C. Traffic Patterns

In order to represent the spatial-temporal characteristics of urban traffic, four synthetic traffic patterns are developed in SUMO simulation (as described in TABLE III):

- In *major/minor road* traffic pattern (P1), the traffic demands of west-east directions (i.e. major road) are higher than north-south directions (i.e. minor road).

- In *through/left-turn lane* traffic pattern (P2), the traffic demands of left-turn lanes of west-east directions are higher than the through lanes.

- In *tidal* traffic pattern (P3), the traffic demands from north and east are higher than the opposite directions.

- In *varying demand* traffic pattern (P4), the traffic demands of through lanes of west-east directions are time-varying, in contrast to the steady traffic demands in the other three traffic patterns.

TABLE III. CONFIGURATIONS OF TRAFFIC PATTERNS

| Traffic pattern | Direction | Arrival rate[a] (vehicles per second) | | |
| --- | --- | --- | --- | --- |
| | | *Through* | *Left-turn* | *Right-turn* |
| P1: Major/minor road | N-S | 0.05 | 0.025 | 0.01 |
| | S-N | 0.05 | 0.025 | 0.01 |
| | E-W | 0.1 | 0.05 | 0.01 |
| | W-E | 0.1 | 0.05 | 0.01 |
| P2: Through/ left-turn lane | N-S | 0.05 | 0.025 | 0.01 |
| | S-N | 0.05 | 0.025 | 0.01 |
| | E-W | 0.05 | 0.1 | 0.01 |
| | W-E | 0.05 | 0.1 | 0.01 |
| P3: Tidal traffic | N-S | 0.1 | 0.08 | 0.01 |
| | S-N | 0.05 | 0.025 | 0.01 |
| | E-W | 0.1 | 0.08 | 0.01 |
| | W-E | 0.05 | 0.025 | 0.01 |
| P4: Varying demand traffic | N-S | 0.05 | 0.025 | 0.01 |
| | S-N | 0.05 | 0.025 | 0.01 |
| | E-W | 0.05 (0-1200 steps) 0.15 (1200-1800 steps) | 0.025 | 0.01 |
| | W-E | 0.15 (0-600 steps) 0.05 (600-1800 steps) | 0.025 | 0.01 |

a. The arriving of vehicles is generated by binomial distribution with specified arrival rates.

### D. Benchmarks

Three traditional traffic signal control methods are used as benchmarks for performance comparison with our RL approach.

- Fixed-time control: signal phase durations are pre-calculated according to the proportion of vehicle arrival rates and fixed during operation.

- Gap-based control: prolong signal phases whenever a continuous (i.e. maximum time gap between successive vehicles ≤ 5s) stream of traffic is detected [18].
- Time loss-based control: prolong signal phases whenever there exists vehicles with accumulated time loss (i.e. $1-v/v_{max}$) exceeding 1s [18].

*E. Results and Discussion*

The performance of our RL approach is evaluated in three areas: (1) training convergence, (2) comparison with benchmarks and (3) generalization across different traffic patterns. Average queue length (i.e. number of halting vehicles per incoming lane) and average wait time (i.e. wait time in second per incoming vehicle) are used as performance criteria.

Fig. 4 shows the training convergence of our RL approach under traffic patterns P1~P3. Traffic pattern P4 is used for testing purpose to validate the generalization ability of our RL approach. At the beginning of the training process, the Q-network explores the control policy by selecting a random action with high probability $\varepsilon$. As training goes on, the Q-network gets positive or negative rewards, depending on whether a correct action has been taken to reduce the number of halting vehicles. The Q-network gradually exploits the control policy and reduces the average queue length and average wait time. Finally, the Q-network achieves a stabilized and converged performance with respect to both average queue length and average wait time.

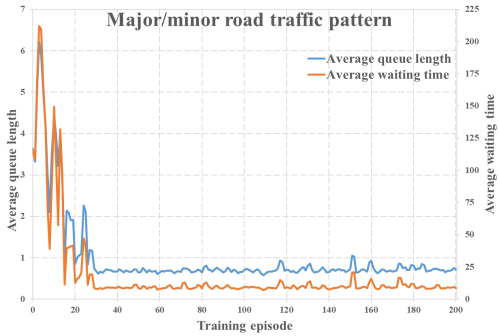

(P1: major/minor road traffic pattern)

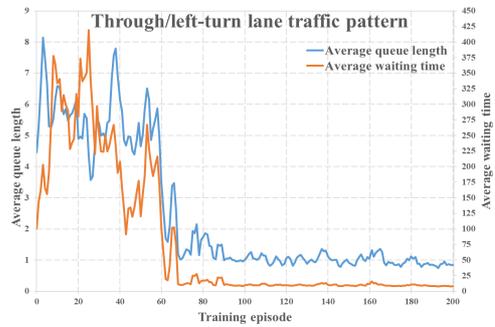

(P2: Through/left-turn lane traffic pattern)

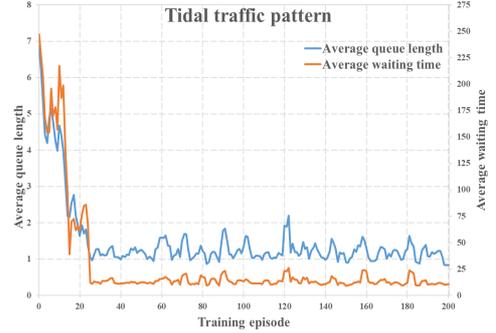

(P3: Tidal traffic pattern)

Figure 4. Convergence of the Q-network (lower is better).

TABLE IV summarizes the evaluations on how the control policy learned by our RL approach generalizes across different traffic patterns. For example, the entry in the P1 row of the P3 column shows the average performance (in 100 attempts) with the RL approach trained in traffic pattern P1 and tested in traffic pattern P3. Overall, our RL approach generalizes well across different traffic patterns with slight performance variations. Especially, the RL approach trained in traffic patterns P1~P3 which feature steady traffic flows also performs well in traffic pattern P4 which has a time-varying traffic flow.

TABLE IV. GENERALIZATION ACROSS TAFFIC PATTERNS

| Train \ Test | P1 | P2 | P3 | P4 |
|---|---|---|---|---|
| P1 | 0.72/9.35[a] | 1.13/11.11 | 1.42/11.68 | 0.56/9.67 |
| P2 | 0.71/10.26 | 0.91/9.09 | 1.04/10.98 | 0.61/8.92 |
| P3 | 0.67/8.98 | 0.96/9.76 | 1.17/12.98 | 0.66/11.30 |

a. average queue length / average waiting tine.

Fig. 5 shows the performance comparisons of our RL approach with the benchmark traffic signal control methods. Box plots in Fig. 5 Are obtained by repeating each method 100 times in each traffic pattern. The bold line in the middle of a box is the median, the lower line of the box is the lower quartile (i.e. 1st quartile or 25th percentile) and the upper line is the upper quartile (i.e. 3rd quartile or 75th percentile). Clearly, our RL approach is able to achieve a better performance in terms of the average queue length and average wait time in each traffic pattern when compared with the benchmarks. Even compared with the second-best benchmark (i.e. time loss-based control), the performance improvements of the RL approach are still significant in all traffic patterns (i.e. ranging from 27%~73% in average queue length and 42%~79% in average wait time).

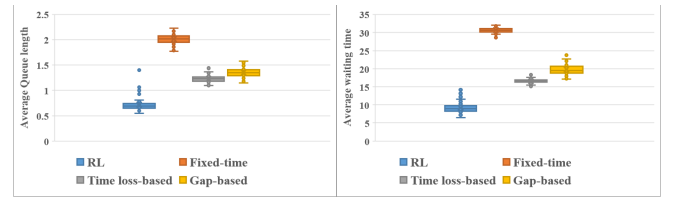

(P1: major/minor road traffic pattern)

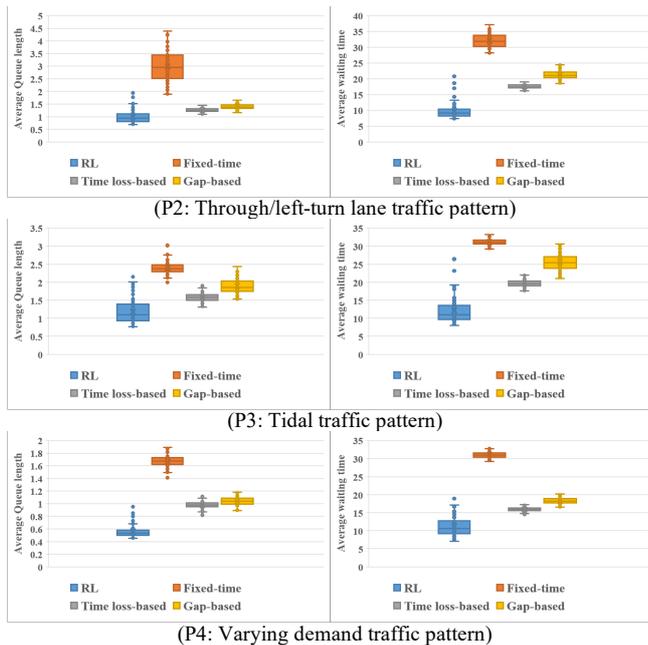

(P2: Through/left-turn lane traffic pattern)

(P3: Tidal traffic pattern)

(P4: Varying demand traffic pattern)

Figure 5. Performance comparisons with benchmarks.

V. CONCLUSION

This paper, we present a neural networked RL approach for intelligent traffic signal control that takes into consideration the spatial-temporal characteristics of urban traffic flows. In order to cope with a large state space which consists of real-time vehicle position and speed, a Q-network is proposed to extract discriminative information from the state space in order to derive an optimal signal control policy. Four traffic patterns are implemented in SUMO simulation. The simulation results demonstrate the superior performance of our RL approach under different traffic patterns against three benchmark methods in terms of queue length and wait time.

There is room for further improvement of the Q-network by tuning its performance in convergence and stability. Advanced techniques such as dueling network [25] and double Q-learning network [26] can be employed. Additionally, extending the RL approach to more complex urban intersection settings such as an arterial or a multi-intersection network presents interesting challenges for exploration of the proposed methodology.

REFERENCES

[1] Y. Wang, X. Yang, H. Liang, and Y. Liu, "A review of the self-adaptive traffic signal control system based on future traffic environment," *J. Adv. Transp.*, p. 12, 2018.
[2] Y. Wang, B. Ning, T. Tang, L. Zhu, and F. R. Yu, "Big data analytics in intelligent transportation systems: A Survey," *IEEE Trans. Intell. Transp. Syst.*, vol. 20, no. 1, pp. 383–398, 2018.
[3] Beijing Transportation Research Center, "Beijing transport annual report," 2017. [Online]. Available: http://www.bjtrc.org.cn/InfoCenter/NewsAttach/2017%E5%B9%B4%E6%8A%A5%E6%9C%80%E7%BB%88.pdf
[4] S. Troia, S. Gao, R. Alvizu, G. A. Maier, and A. Pattavina, "Identification of tidal-traffic patterns in metro-area mobile networks via Matrix Factorization based model," *2017 IEEE Int. Conf. Pervasive Comput. Commun. Work.* 2017, pp. 297–301, 2017.
[5] Y. T. Wu and C. H. Ho, "The development of Taiwan arterial traffic-adaptive signal control system and its field test: A Taiwan experience," *J. Adv. Transp.*, vol. 43, no. 4, pp. 455–480, 2009.
[6] U.S. Department of Transportation, "Traffic signal timing manual," 2017. [Online]. Available: https://ops.fhwa.dot.gov/publications/fhwahop08024/index.htm#toc
[7] S. Göttlich, M. Herty, and U. Ziegler, "Modeling and optimizing traffic light settings in road networks," *Comput. Oper. Res.*, vol. 55, pp. 36–51, 2015.
[8] I. Kosonen, "Multi-agent fuzzy signal control based on real-time simulation," *Transp. Res. Part C Emerg. Technol.*, vol. 11, no. 5, pp. 389–403, 2003.
[9] Y. Liu, L. Liu, and W. P. Chen, "Intelligent traffic light control using distributed multi-agent Q-learning," in *Proc. 21th Int. IEEE Conf. Intell. Transp. Syst. (ITSC)*, 2018, vol. 2018–March, pp. 1–8.
[10] S. El-Tantawy, B. Abdulhai, and H. Abdelgawad, "Design of reinforcement learning parameters for seamless application of adaptive traffic signal control," *J. Intell. Transp. Syst. Technol. Planning, Oper.*, vol. 18, no. 3, pp. 227–245, 2014.
[11] Prabuchandran K.J., Hemanth Kumar A.N, and S. Bhatnagar, "Multi-agent reinforcement learning for traffic signal control," in *Proc. 17th Int. IEEE Conf. Intell. Transp. Syst. (ITSC)*, 2014, pp. 2529–2534.
[12] H. Wei, H. Yao, G. Zheng, and Z. Li, "IntelliLight: A reinforcement rearing approach for intelligent traffic light control," in *KDD2018 – Proc. 24th ACM SIGKDD Int. Conf. Knowl. Discovery and Data Mining*, 2018, pp. 2496–2505.
[13] L. Li, Y. Lv, and F. Wang, "Traffic signal timing via deep reinforcement Learning," *IEEE/CAA J. Autom. Sin.*, vol. 3, no. 3, pp. 247–254, 2016.
[14] S. M. A. Shabestary and B. Abdulhai, "Deep learning vs. discrete reinforcement learning for adaptive traffic signal control," in *Proc. 21th Int. IEEE Conf. Intell. Transp. Syst. (ITSC)*, 2018, vol. 2018–Nov., pp. 286–293.
[15] J. Zeng, J. Hu, and Y. Zhang, "Adaptive traffic signal control with deep recurrent Q-learning," in *Proc. IEEE Intell. Veh. Symp. (IV)*, 2018, pp. 1215–1220.
[16] X. Liang, X. Du, G. Wang, and Z. Han, "Deep reinforcement learning for traffic light control in vehicular networks," *IEEE Trans. Veh. Technol.*, vol. XX, no. XX, 2018.
[17] V. Mnih, K. Kavukcuoglu, D. Silver, A. A. Rusu, J. Veness, M. G. Bellemare, A. Graves, M. Riedmiller, A. K. Fidjeland, G. Ostrovski, S. Petersen, C. Beattie, A. Sadik, I. Antonoglou, H. King, D. Kumaran, D. Wierstra, S. Legg, and D. Hassabis, "Human-level control through deep reinforcement learning," *Nature*, vol. 518, p. 13, Feb. 2015.
[18] D. Krajzewicz, J. Erdmann, M. Behrisch, and L. Bieker, "Recent development and applications of SUMO-Simulation of Urban MObility," *Int. J. Adv. Syst. Meas.*, vol. 5, no. 3, pp. 128–138, 2012.T.
[19] Schaul, J. Quan, I. Antonoglou, and D. Silver, "Prioritized experience replay," in *ICLR*, 2016, pp. 1–21.
[20] C. J. C. H. Watkins and P. Dayan, "Q-learning," *Mach. Learn.*, vol. 8, no. 3–4, pp. 279–292, 1992.
[21] R. S. Sutton and A. G. Barto, "Reinforcement learning: An introduction," *The MIT Press*, 2018.
[22] S. Krauß, "Towards a unified view of microscopic traffic flow theories," *IFAC Trans. Syst.*, vol. 30, no. 8, pp. 901–905, 1997.
[23] A. Paszke, S. Gross, S. Chintala, and G. Chanan, "PyTorch: Tensors and dynamic neural networks in Python with strong GPU acceleration," 2017. [Online]. Available: http://pytorch.org/
[24] R. Girshick, "Fast R-CNN," In *ICCV*, pp. 1440-1448. 2015.
[25] Z. Wang, T. Schaul, M. Hessel, H. Van Hasselt, M. Lanctot, and N. De Freitas, "Dueling network architectures for deep reinforcement learning," in *Proc. 33rd Int. Conf Int. Conf. Mach. Learn.*, 2016, pp. 1995–2003.
[26] H. Van Hasselt, "Double Q-learning," in *NIPS Proc.*, 2010, pp. 1–9.